\documentstyle[aps,preprint]{revtex}
\def\be{\begin{equation}}
\def\ee{\end{equation}}
\def\bea{\begin{eqnarray}}
\def\eea{\end{eqnarray}}
\begin{document}
\begin{center}
{\LARGE \bf Comment on ``Quantum bound \\
\vskip.3cm
states with zero binding energy''}\\
\vskip1cm
{\bf Sergio A. Hojman$^\ast$ and Dar\'\i o N\'u\~nez$^\dagger$}\\
\vskip5mm
$^{\ast}$ Departamento de F\'\i sica, Facultad de Ciencias, \\
Universidad de Chile, Casilla 653, Santiago, Chile \\
\vskip3mm
$^{\dagger}$ Instituto de Ciencias Nucleares, UNAM, \\Circuito Exterior, C.U.,
A.P. 70-543, M\'exico, D. F. 04510,  M\'exico.
\end{center}

\begin{abstract}
The purpose of this Comment is to show that the solutions to the zero energy
Schr\"odinger equations for monomial central potentials discussed in a
recently published Letter, may also be obtained from the corresponding
free particle solutions in a straight forwardly way, using an algorithm
previously devised by us.
New solutions to the zero energy Schr\"odinger equation are also exhibited.
\end{abstract}
\vspace{7mm}

\vfill
\eject

Looking for relations between the solutions of otherwise seemingly unrelated
problems has been a task undertaken by several researchers at many different
times and places. For instance, in recent years, Kustaanheimo and Stiefel
\cite{KS} found a remarkable transformation which relates the solutions of
the harmonic oscillator to the ones of the hydrogen atom both in classical
and in quantum mechanics. Another interesting example is provided by the 1981
Witten's rediscovery \cite{Witten} of a
transformation obtained by Darboux in 1882 \cite{Darboux}, which links the
solutions of the one-dimensional Schr\"odinger equation of two
super--symmetrically related potentials.  There are, of course, many other
transformations which establish connections between apparently different
problems of mathematical physics, but we will not comment on them any
further.

In 1991 we devised, in collaboration with S. Chayet and M. Roque \cite{HN1},
an algorithm which may be considered as a generalization of the
Kustaanheimo and Stiefel transformation.
We were able to explicitly construct transformations to relate the solutions
of different and seemingly unrelated dynamical systems. Amazingly enough,
{\it the same transformation} works equally well for both the classical and
quantum versions of the problems under consideration. The algorithm allowed
us to solve cases that were considered not solvable up to then. These
transformations may also be applied to many other physical systems, including
field theoretical examples. We also showed that problems
in geometrical and wave optics may be considered as particular cases of
problems in classical and quantum mechanics, respectively. In fact, to
deal with optics, it is enough to consider just one of the many possible
energies of the system, as we explicitly showed in our article, using a
simple relation which transforms potentials into refraction indices, and
{\it vice--versa}. Due to this simplification, our results in optics turn out
to be much stronger than the ones obtained in other cases, allowing us to
relate, and solve, a large class of refraction indices and, due to the
complete analogy between wave optics and the Schr\"odinger equation for the
case of zero energy, it is possible to solve this equation for several
families of potentials. Each one of these families possesses infinitely many
members.

In a recently published Letter \cite{Nieto}, Daboul and Nieto
solved the $E=0$ Schr\"odinger equation for central potentials of the form
$V=-{{|\gamma|}r^{-\nu}}$, for arbitrary $\nu$, and showed that these
solutions have several interesting and unusual physical properties.

In this Comment, we show that their results may be straight forwardly obtained
by the use of our algorithm for the particular case of the zero energy
Schr\"odinger equation.  We give a brief review of the pertinent part of our
algorithm and reproduce the
solutions of the $E=0$ Schr\"odinger
equation for infinitely many potentials which are related to the free
particle. The $V=-{{|\gamma|}r^{-\nu}}$ case is obtained as a special case of
the potentials related to the free particle. We also present a few
generalizations of the cases mentioned above.

Consider the two-dimensional Schr\"odinger
equation \cite{HN1} (generalizations to higher dimensions wil be dealt with
at the end of this Comment). The $E=0$ case turns out to be very simple.
In fact,
consider the $E=0$ Schr\"odinger equation associated with an arbitrary
potential, which in terms of complex coordinates $z$ and $z^*$ reads
\be
\left( -{1\over 8}\partial_z\,\partial_{z^*} +
V(z,z^*)\right)\psi(z,z^*)=0.
\label{scheq}
\ee
where $z=x+iy$. The analytic transformation,
\be
z=f(\omega),\,\,\, \omega=u+iv, \label{ctrc}
\ee
{\it i.e.},
\be
x=x(u,v),\,\,y=y(u,v),
\ee
such that
\be
{{\partial x}\over{\partial u}}={{\partial y}\over{\partial v}}\, , \,\,\,\,\,
\,{{\partial x}\over{\partial v}}=-{{\partial y}\over{\partial u}}\, ,
\ee
changes the
$E=0$ Schr\"odinger equation into itself for the new potential
\be
U=\left|{{d f}\over{d \omega}}\right|^2\,V(f(\omega),f^*(\omega^*))\,\, ,
\ee
{\it i.e.},
\be
\left( -{1\over 8}\partial_\omega\,\partial_{\omega^*} +
U(\omega,\omega^*)\right)\phi(\omega,\omega^*)=0.
\label{scheqt}
\ee

Its solution, in terms of the one for (\ref{scheq}) is
\be
\phi(\omega,\omega^*)=\psi(f(\omega),f^*(\omega^*)).
\ee

The two-dimensional $E=0$ Schr\"odinger equation for a free particle,
in polar coordinates, is
\be
\left( r^2\,{{\partial^2}\over{\partial r^2}}\,+
r\,{{\partial}\over{\partial r}}\, +
{{\partial^2}\over{\partial \theta^2}}\,+ \,r^2\, \right)\,
\Psi(r,\theta)= 0.\label{eq:fp2}
\ee
where we have chosen the free particle potential $V=-(1/2)$ for convenience.

Its solutions are \cite{Kamke}
\be
\Psi(r,\theta)=J_{L_d}(r)\,e^{i\,L_d\,\theta}, \label{soli}
\ee
where $J_{L_d}(r)$ are Bessel functions and $L_d$ is a constant
related with the conservation of angular momentum.

Now let us make a conformal transformation, $(r,\,\theta)\to (\rho, \phi)$.
Recall that the Cauchy-Riemann conditions written in terms of polar coordinates
are
\be
{{\partial \rho}\over{\partial r}}=
{\rho \over r}\,{{\partial \phi}\over{\partial \theta}}\, , \,\,\,\,\,\,\,\,\,
{{\partial \rho}\over{\partial \theta}}=-r\,\rho\,
{{\partial \phi}\over{\partial r}}\, .\label{confp}
\ee
The $E=0$ Schr\"odinger equation for the free particle, Eq.~(\ref{eq:fp2}),
now becomes
\be
\left( \rho^2\,{{\partial^2}\over{\partial \rho^2}}\,+
\rho\,{{\partial}\over{\partial \rho}}\, +
{{\partial^2}\over{\partial \phi^2}}\,- 2\,\rho^2\,U(\rho, \phi) \right)\,
\Phi(\rho,\phi)= 0,\label{eq:tr}
\ee
where the potential $U(\rho, \phi)$ is given by
\be
U(\rho, \phi)=\,-\,{1\over2}{ {r^2(\rho, \phi)} \over { {r^2(\rho, \phi)\,
({{\partial \rho}\over{\partial r}})^2 +
({{\partial \rho}\over{\partial \theta}})^2}}}.\label{eq:pu}
\ee

Note that this result is valid for any analytical transformation.
To recover the potential $U=-(1/2)\,\sigma^2\,\rho^{(2\,\alpha -2)}$,
dealt with in Ref. \cite{Nieto}, just
perform the conformal transformation
\be
z={\sigma\over \alpha}\,\omega^\alpha,
\label{transfc}
\ee
or, in real polar variables language,
\be
r={\sigma\over \alpha}\,\rho^\alpha,\,\,\theta=\alpha\,\phi,
\label{transf}
\ee
with $\sigma$ and $\alpha$ non--vanishing arbitrary constants, and
choose $L_d=l/\alpha$. We have thus showed that a very particular conformal
transformation allows us to recover the solutions of Ref. \cite{Nieto},
starting from the free particle problem. The most general potential
attainable starting from the free particle equation is already quoted
in Eq.~(\ref{eq:pu}).

The d--dimensional Schr\"odinger equation is easily obtained, starting from
equation (\ref{eq:tr}) by defining
the function $\Xi(\rho,\phi)$
\be
\Xi(\rho,\phi)=\rho^{{2-d}\over 2}\Phi(\rho, \phi),
\ee
to get
\be
\left( \rho^2\,{{\partial^2}\over{\partial \rho^2}}\,+
(d-1)\rho\,{{\partial}\over{\partial \rho}}\, +
{{\partial^2}\over{\partial \phi^2}}\,- 2\,\rho^2\,\tilde{U}(\rho, \phi)
\right)\,\Xi(\rho,\phi) = 0,\label{eq:tr2}
\ee
which is the $E=0$ d-dimensional Schr\"odinger equation, with symmetry in
$(d-2)$ coordinates, for the potential $\tilde U(\rho, \phi)$ given by
\be
\tilde U(\rho, \phi)=U(\rho, \phi) -{{(d-2)^2}\over {8 \rho^2}}.
\label{eq:puf}
\ee
Its solutions are given by
\be
\Xi(\rho, \phi)=\rho^{{2-d}\over 2}\,J_{L_d}\big( r(\rho, \phi) \big)\,
e^{i\,L_d\,\theta(\rho, \phi)},
\ee
which, as it was already mentioned, is the solution to the zero energy
Schr\"o\-dinger equation for any potential of the form given by
Eqs.~(\ref{eq:pu}) and (\ref{eq:puf}).

We now briefly outline how to enlarge the class of potentials solved
above, by using the algorithm devised in \cite{HN1}. One
way to do this is by starting from a solvable $E=0$ central problem not
related to the $E=0$ free particle equation and use the tranformation
given by Eq. (\ref{transf}). This algorithm will produce new d-dimensional
$E=0$ central potential solvable problems. For instance, if we start from the
central two-dimensional $E=0$ problem defined by the potential
\be
V= a\,r^2\,+\,b\,r^{-2}\,+\,c\,= a\,z\,z^*\,+\,\frac{b}{z\,z^*}\,+\,c\,\,\,\,,
\label{np1}
\ee
and use the transformation given by Eqs. (\ref{transfc}) or (\ref{transf})
the new potential is
\be
U= \frac{a\,\sigma^4}{\alpha^2}\,\rho^{4\,\alpha\,-\,2}\,+\,b\,\alpha^2\,
\rho^{-2}\,+\,c\,\sigma^2\,\rho^{2\,\alpha\,-\,2}\,\,\,\,,
\label{np2}
\ee
If we are interested in two-dimensional problems only, then one can use
{\it any} conformal transformation, not just Eq. (\ref{transf}) and  solve
central and non--central potential problems, starting from the $E=0$ free
particle (or other solvable potential) Schr\"odinger equation. The
transformation $z=e^\omega$, relates the zero energy free particle system to
the same kind of problem defined by the transformed potential
$U=-(1/2)e^{2u}$, for instance.

In summary, we have made use of a previously devised algorithm \cite{HN1} to
show how to
obtain the solutions to the $E=0$ Schr\"odinger equation for the cases
studied in \cite{Nieto} starting from the free particle solutions. We have
furthermore extended those results to include other families of potentials
which can be easily dealt with using the aforementioned procedure. Other
cases including examples in field theory will be considered in forthcoming
articles.

\centerline{\bf Acknowledgements}

This work was partially supported by {\bf C}entro para la
{\bf I}nvestigaci\'on y la {\bf EN}se\~nanza de la {\bf CI}encia y sus
{\bf A}plicaciones ({\bf CIENCIA}, Chile), grant 93--0883 (FONDECYT, Chile)
and a bi--national grant
funded by Comisi\'on Nacional de Investigaci\'on Cient\'\i fica y
Tecnol\'ogica--Fundaci\'on Andes (Chile) and Consejo Nacional de Ciencia y
Tecnolog\'\i a (M\'exico). S. A. H. would like to express his deep gratitude
to the Instituto de Ciencias Nucleares, Universidad Nacional Aut\'onoma de
M\'exico for warm hospitality while this work was done.



\end{document}